\documentclass[footinbib,aps,prl,reprint,superscriptaddress,nopacs]{revtex4-1}

\usepackage{amsmath,amssymb,braket,upgreek,bm,verbatim}
\usepackage[absolute]{textpos}
\usepackage{physics, graphicx, tabularx,braket}
\usepackage{xcolor}

\usepackage[absolute]{textpos}
\usepackage{multirow}

\definecolor{darkblue}{rgb}{0,0,0.5}

\DeclareMathOperator{\sF}{\mathcal{F}}

\usepackage{hyperref}
\hypersetup{
colorlinks=true,
linkcolor=black,
filecolor=blue,
citecolor=darkblue,  
urlcolor=black,
}

\definecolor{joeBlue}{RGB}{0,163,224}

\begin{document}
\title{Procrustean entanglement concentration in quantum--classical networking}

\author{Hsuan-Hao Lu}
\email{luh2@ornl.gov}
\author{Muneer~Alshowkan}
\affiliation{Quantum Information Science Section, Computational Sciences and Engineering Division, Oak Ridge National Laboratory, Oak Ridge, Tennessee 37831, USA}

\author{Jude Alnas}
\affiliation{School of Electrical and Computer Engineering, Duke University, Durham, North Carolina 27708, USA}

\author{Joseph M. Lukens}
\affiliation{Quantum Information Science Section, Computational Sciences and Engineering Division, Oak Ridge National Laboratory, Oak Ridge, Tennessee 37831, USA}
\affiliation{Research Technology Office and Quantum Collaborative, Arizona State University, Tempe, Arizona 85287, USA}

\author{Nicholas~A.~Peters}
\affiliation{Quantum Information Science Section, Computational Sciences and Engineering Division, Oak Ridge National Laboratory, Oak Ridge, Tennessee 37831, USA}

\date{\today}

\begin{textblock}{13.3}(1.4,15)
\noindent\fontsize{7}{7}\selectfont \textcolor{black!30}{This manuscript has been co-authored by UT-Battelle, LLC, under contract DE-AC05-00OR22725 with the US Department of Energy (DOE). The US government retains and the publisher, by accepting the article for publication, acknowledges that the US government retains a nonexclusive, paid-up, irrevocable, worldwide license to publish or reproduce the published form of this manuscript, or allow others to do so, for US government purposes. DOE will provide public access to these results of federally sponsored research in accordance with the DOE Public Access Plan (http://energy.gov/downloads/doe-public-access-plan).}
\end{textblock}

\begin{abstract}
The success of a future quantum internet will rest in part on the ability of quantum and classical signals to coexist in the same optical fiber infrastructure, a challenging endeavor given the orders of magnitude differences in flux of single-photon-level quantum fields and bright classical traffic. We theoretically describe and experimentally implement Procrustean entanglement concentration for polarization-entangled states  contaminated with classical light, showing significant mitigation of crosstalk noise in dense wavelength-division multiplexing. Our approach leverages a pair of polarization-dependent loss emulators to attenuate highly polarized crosstalk that results from imperfect isolation of conventional signals copropagating on shared fiber links. We demonstrate our technique both on the tabletop and over a deployed quantum local area network, finding a substantial improvement of two-qubit entangled state fidelity from approximately 75\% to over 92\%. This local filtering technique could be used as a preliminary step to reduce asymmetric errors, potentially improving the overall efficiency when combined with more complex error mitigation techniques in future quantum repeater networks.
\end{abstract}

\maketitle

\textit{Introduction.---}Quantum-classical coexistence within fiber-optic resources will streamline the implementation of quantum networks, enabling efficient utilization of infrastructure and reducing deployment costs. One promising pathway relies on distinguishing classical and quantum channels by wavelength. Although originally pursued through coarse wavelength-division multiplexing (CWDM)
~\cite{Townsend1997, Patel2012, Huang2015, Mao2018, Burenkov2023, Thomas2023}, quantum--classical coexistence efforts are increasingly targeting dense wavelength-division multiplexing (DWDM) with channels of 200~GHz or less, for even greater spectral efficiency~\cite{Peters2009, Eraerds2010, Choi2010, Qi2010, Patel2014, Chapman2023a, Chapman2023b, Honz2023}. Given the stark contrast in brightness between classical and quantum signals, such ultratight spacings are highly susceptible to imperfect filter isolation and unwanted noise infiltrating the quantum output~\cite{Peters2009, Eraerds2010, Chapman2023a, Honz2023}. 
At whatever level of crosstalk is present, the strong temporal correlations between entangled photons can be leveraged to some extent for filtering uncorrelated noise by reducing the coincidence detection window. Nevertheless, the timing jitter between two nodes ultimately determines how narrow this window can be before genuine coincidences are lost, a limitation becoming especially pronounced in geographically separated nodes~\cite{Alshowkan2022a, Burenkov2023}.

In this Letter, we introduce a crosstalk mitigation method tailored to polarization-entangled photons tainted by highly polarized classical signals, a practical source of error from typically polarized conventional communications signals. For appropriate orientations, the noisy density matrix maps approximately to the class of
maximally entangled mixed states (MEMS)~\cite{Munro2001, Wei2003, Peters2004a}, whose entanglement can be concentrated through local ``Procrustean'' filtering~\cite{Bennett1996, Kwiat2001, Peters2004a, Wang2006}.
Additionally, our approach resembles the polarization dependent loss (PDL) compensation possible for Bell states~\cite{Jones2018, Kirby2019, Riccardi2021}, whereby reductions in concurrence due to PDL on one photon can be completely compensated (in postselection) by applying PDL 
to the other. 
This situation facilitates the filtering of noisy quantum states while retaining maximal entanglement in the Bell state portion and suppressing nonentanlged terms. After simulating a general model, we experimentally demonstrate the approach with programmable polarization-dependent loss emulators (PDLEs; OZ Optics) applied to entangled photons contaminated with crosstalk from copropragating lasers. In all cases examined---both on the optical table and over a deployed network---fidelity is found to increase in good agreement with theory. Overall, our method provides an additional error-mitigation layer for quantum--classical coexistence networks that  can be applied, e.g., for dynamic removal of asymmetric noise.

\begin{figure}[tb!]
\centering
\includegraphics[width=\linewidth]{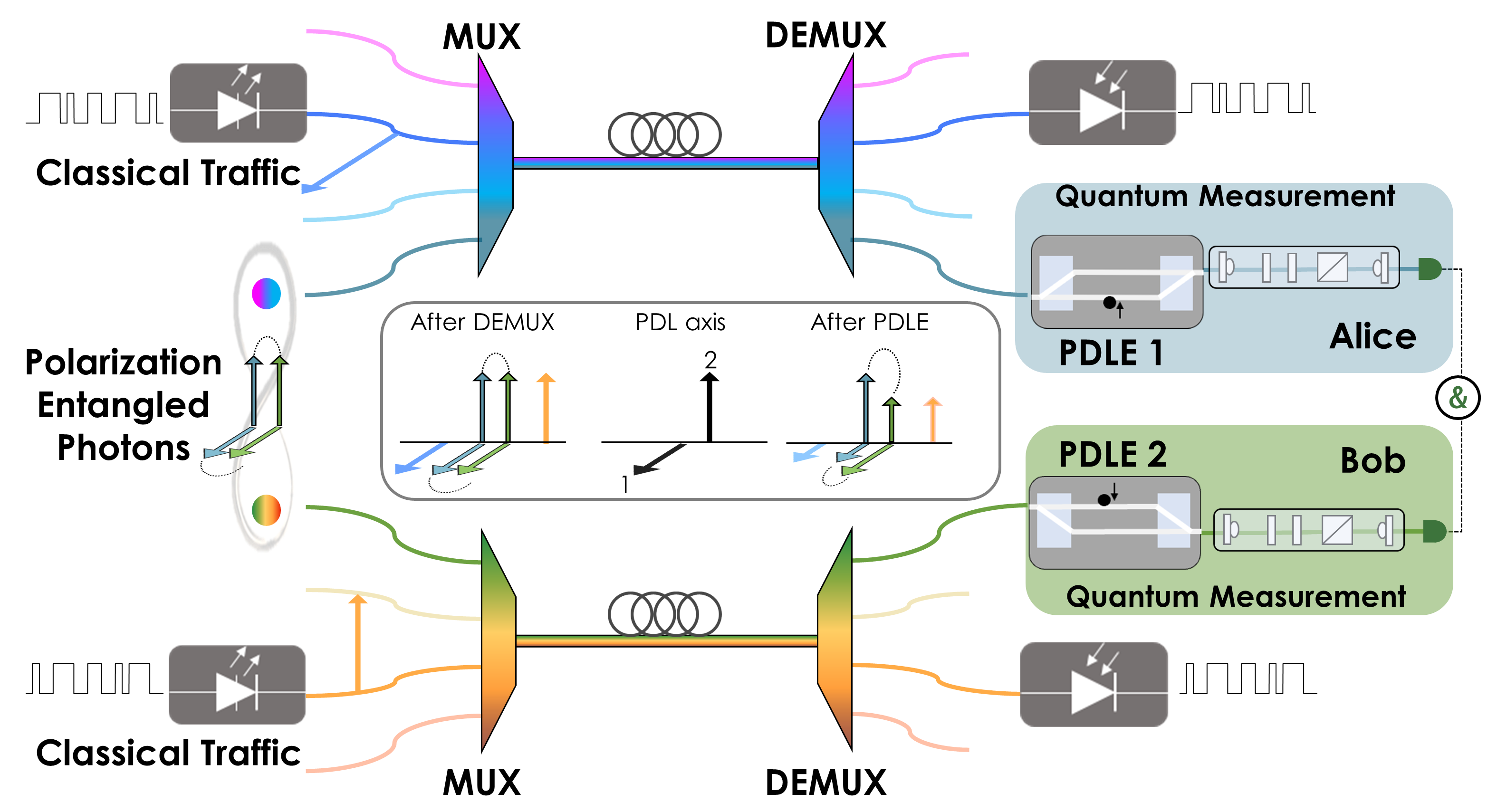}
\caption{Concept of Procrustean filtering for polarization-entangled photons. Crosstalk suppression can be achieved by aligning the PDLE axes and classical traffic polarizations accordingly.}
\label{fig1}
\end{figure}

\textit{Motivation.---}Figure~\ref{fig1} illustrates the proposed scheme. Consider an ideal Bell state $\ket{\Phi^+}=\frac{1}{\sqrt{2}}(\ket{HH}+\ket{VV})$ as the input, where $\ket{H}$ ($\ket{V}$) denotes a horizontally (vertically) polarized state. Following generation, signal and idler photons are physically separated into two optical arms destined for Alice ($A$) and Bob ($B$), each combined with highly polarized classical traffic through a DWDM multiplexer (MUX). Upon reaching the receivers, photons are wavelength-demultiplexed using another DWDM demultiplexer (DEMUX). Due to imperfect spectral filtering and channel isolation, the quantum channel now includes a small portion of classical crosstalk noise. Alice and Bob send their respective signals through a PDLE module designed to controllably replicate PDL in optical links~\cite{Gisin1995, Mecozzi2002}. The PDLE module spatially separates the input into two orthogonal polarizations, applying user-defined attenuation to one of them, referred to as the PDL axis, while passing the other unaltered. The two polarizations are then recombined into a single fiber-optic spatial mode at the output.

To optimize the noise suppression possible with PDLEs
, intuition suggests orienting the PDL axes orthogonally ($H/V$) with respect to the initial reference frame to ensure equal attenuation of the terms of interest in $\ket{\Psi^+}$, while aligning the classical traffic to these axes for greatest suppression. In this scenario, a reasonable approximation of the initial noisy quantum state (before any PDL) is $\rho \approx \gamma\ket{\Phi^+}\bra{\Phi^+} + (1-\gamma)\ket{HV}\bra{HV}$, where $\gamma\in[0,1]$ defines the relative weight of accidentals stemming from the noise photons in comparison to the genuine coincidences between the entangled pairs. 
For $\gamma\in[2/3, 1]$ this state is precisely that of the MEMS~I subclass~\cite{Munro2001,Wei2003,Peters2004a}, which is a sufficient (though not necessary) criterion for 
our concentration method. 

With the introduction of two PDLEs imparting variable transmissivities $T_H,T_V\in[0,1]$ along their respective PDL axes, the output density matrix becomes $\rho \propto \gamma\left(\sqrt{T_H}\ket{HH} + \sqrt{T_V}\ket{VV} \right) \left(\sqrt{T_H}\bra{HH} + \sqrt{T_V}\bra{VV}\right) + (1-\gamma)T_H T_V\ket{HV}\bra{HV}$. Notably, the unwanted term $\ket{HV}\bra{HV}$ decreases more rapidly than the rest of the matrix entries as $T_H$ and $T_V$ decrease. In particular, the fidelity with respect to the maximally entangled ideal ($\sF=\bra{\Phi^+}\rho\ket{\Phi^+}$) approaches unity as $T_H=T_V \rightarrow 0$, implying an arbitrary increase (at the expense of flux). Now, this model is only approximate in that it neglects other noise terms, such as coincidences \emph{between} crosstalk and input photons, but it summarizes the basic concept. In the following, we introduce the more complete model accounting for all accidental contributions in the system of interest.

\textit{Model.---}Consider polarization-entangled photons generated in the state $\ket{\Phi^+}$ at rate $\mu$. Subsequently, the idler (signal) photon is separated into optical arm $A$ ($B$) with pathway efficiency $\eta_A$ ($\eta_B$) encompassing all loss effects from generation through detection. The rate of photon detection from classical crosstalk is $\nu_A$ ($\nu_B$) in arm $A$ ($B$) when the polarization analyzer is aligned to $H$ ($V$) and $T_H=1$ ($T_V=1$). 
Additionally, we consider dark counts at detector $A$ ($B$) with rate $d_A$ ($d_B$). The coincidence window is set to $\tau$, with the assumption that this window is significantly longer than the two-photon temporal correlations.

\begin{figure*}[tb!]
\includegraphics[width=\linewidth,trim=1cm 0cm 1cm 0cm,clip]{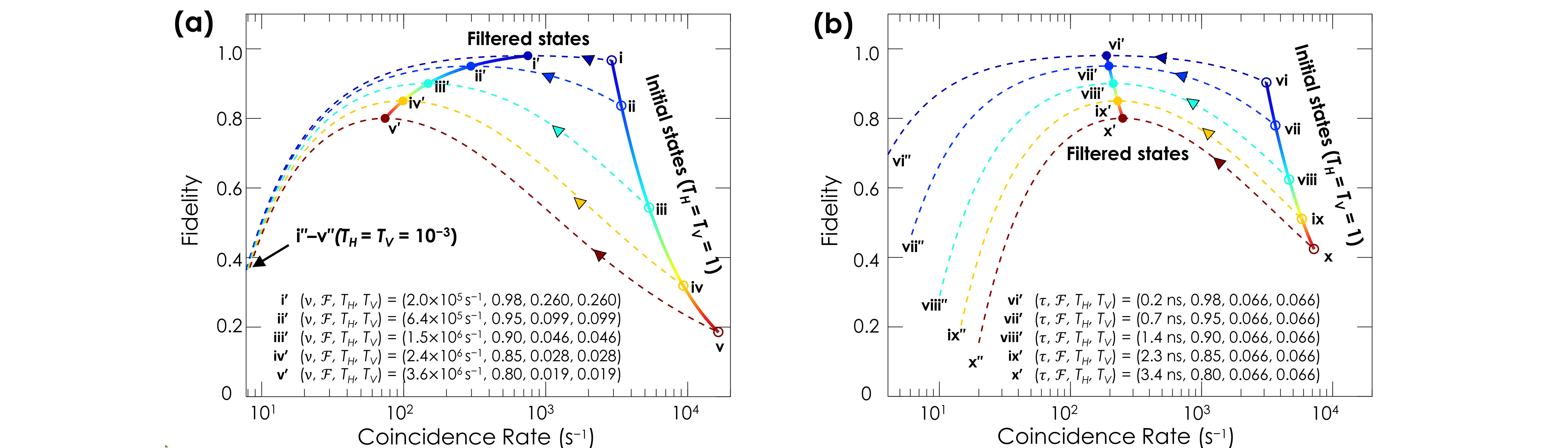}
\caption{Numerical simulation of Procrustean filtering with fixed parameters $\{\mu, \eta_A, \eta_B, d_A, d_B\}=\{7\times10^6, 0.02, 0.02, 100, 100\}$. (a) Variable noise rates $\nu_A =\nu_B =\nu\in [2\times10^5, 3.6\times10^6]$ with a coincidence window $\tau=10^{-9}$. (b) Variable coincidence window $\tau\in [2\times10^{-10}, 3.4\times10^{-9}]$ with a noise rate of $\nu_A =\nu_B = 10^6$. Solid curves represent initial noisy states (right) and optimized filtered states (left). Dashed curves illustrate state evolution as $T_H = T_V$ is scanned from $1$ to $10^{-3}$.}
\label{fig2}
\end{figure*}

Under these conditions, the complete received quantum state can be written as 
\begin{widetext}
\begin{multline}
\label{eq:theory}
\rho = \frac{1}{R} \Bigg\{ \frac{\mu\eta_A\eta_B}{2} \left(\sqrt{T_H}\ket{HH} + \sqrt{T_V}\ket{VV} \right) \left(\sqrt{T_H}\bra{HH} + \sqrt{T_V}\bra{VV}\right) \\ 
+ \tau \left[\frac{\mu\eta_A}{2} \left( T_H\ket{H}\bra{H} +  \ket{V}\bra{V} \right) + \nu_A T_H \ket{H}\bra{H} + d_A I_2\right] \\ \otimes \left[\frac{\mu\eta_B}{2} \left( \ket{H}\bra{H} +  T_V\ket{V}\bra{V} \right) + \nu_B T_V \ket{V}\bra{V} +  d_B I_2\right] \Bigg\}, 
\end{multline}
\end{widetext}
where $I_2$ is the $2\times 2$ identity and $R$ denotes the coincidence rate over all outcomes in a basis, such that $\Tr\rho=1$.
The quantum state contains a correlated term from photons in the same pair (scaled by $\mu\eta_A\eta_B$) and an uncorrelated term (scaled by $\tau$), which can be derived following standard arguments for the rate of accidentals as the product of the rates on the individual detectors multiplied by $\tau$~\cite{Eckart1938, Pearson2010}---valid in the limit of small detection probabilities. 
The noise term thereby consists of a product of the marginal states at each detector, in brackets $[\cdot]$ in Eq.~(\ref{eq:theory}), each of which contains three contributions: (i)~partially polarized light from the filtered entangled photons (scaled by $\mu\eta_{A(B)}$), (ii)~attenuated linearly polarized light from classical crosstalk ($\nu_{A(B)}$), and (iii)~white noise from detector dark counts ($d_{A(B)}$). 

Due to the physical effects reflected in (i) and (iii), Eq.~(\ref{eq:theory}) no longer matches a true MEMS~I form, yet the classical crosstalk term (proportional to $\nu_A\nu_B$) can still be selectively suppressed relative to the correlated term. 
If $T_H=T_V=T$, for example, the flux from the former drops quadratically ($T^2$), while the latter only linearly ($T$). 
As $T$ decreases, at some point the multipair (i) and dark count effects (iii) will take over; e.g., the extreme case of $T_H=T_V =0$ would completely suppress the crosstalk but lead to a separable state
. Consequently, for any state of the form in Eq.~(\ref{eq:theory}), there exists an optimal pair of PDLE filters $(T_H, T_V)$ to maximize $\sF$.

In Fig.~\ref{fig2}(a), we model these tradeoffs in a scenario with 
$\mu=6\times10^6$~s$^{-1}$, 
$\eta_A=\eta_B=0.02$, 
$d_A=d_B=100$~s$^{-1}$, and 
$\tau =1$~ns. We introduce variable noise in both arms, with $\nu_A =\nu_B =\nu\in[2,36]\times 10^5$~s$^{-1}$. The five points (Roman numerals i--v) in Fig.~\ref{fig2}(a) correspond to distinct crosstalk levels $\nu$ with no filtering applied ($T_H=T_V=1$) and represent initial noisy quantum states. For each of these cases, we numerically 
find the $(T_H, T_V)$ pair that maximizes $\sF$, which are are shown as primed values (i$'$--v$'$) and reach $\sF\in\{0.98, 0.95, 0.90, 0.85, 0.80\}$. Our results consistently indicate the optimal filtering occurs when $T_H=T_V$
, aligning with the case for ideal MEMS~I states. We also illustrate the state evolution concerning PDLE filtering by mapping the fidelity/rate trajectories for $T_H=T_V=T\in[0.001,1]$, 
as represented by the dashed curves. 
The maximum attainable fidelity, along with the optimal filtering levels, depends on the noisiness of the initial states. 
Beyond this threshold, $\sF$ decreases as the multipair effect (the unfiltered $\ket{VH}\bra{VH}$ term) increasingly dominates.

Likewise, the influence of crosstalk varies as the coincidence window $\tau$ changes. In Fig.~\ref{fig2}(b), we consider a scenario where photon flux and background levels are fixed at $(\mu, \eta_A, \eta_B, \nu_A, \nu_B, d_A, d_B)=(7\times10^6~\mathrm{s}^{-1}, 0.02, 0.02, 10^6\,\mathrm{s}^{-1}, 10^6\,\mathrm{s}^{-1}, 100\,\mathrm{s}^{-1}, 100\,\mathrm{s}^{-1})$, while $\tau$ varies in $[0.2, 3.4]$~ns. 
As expected, the optimal fidelity is contingent upon the noisiness of the original state. Nevertheless, the optimal filtering configuration ($T_H,T_V$) remains consistent across all cases. 

\begin{figure}[t!]
\centering
\includegraphics[width=3.5in]{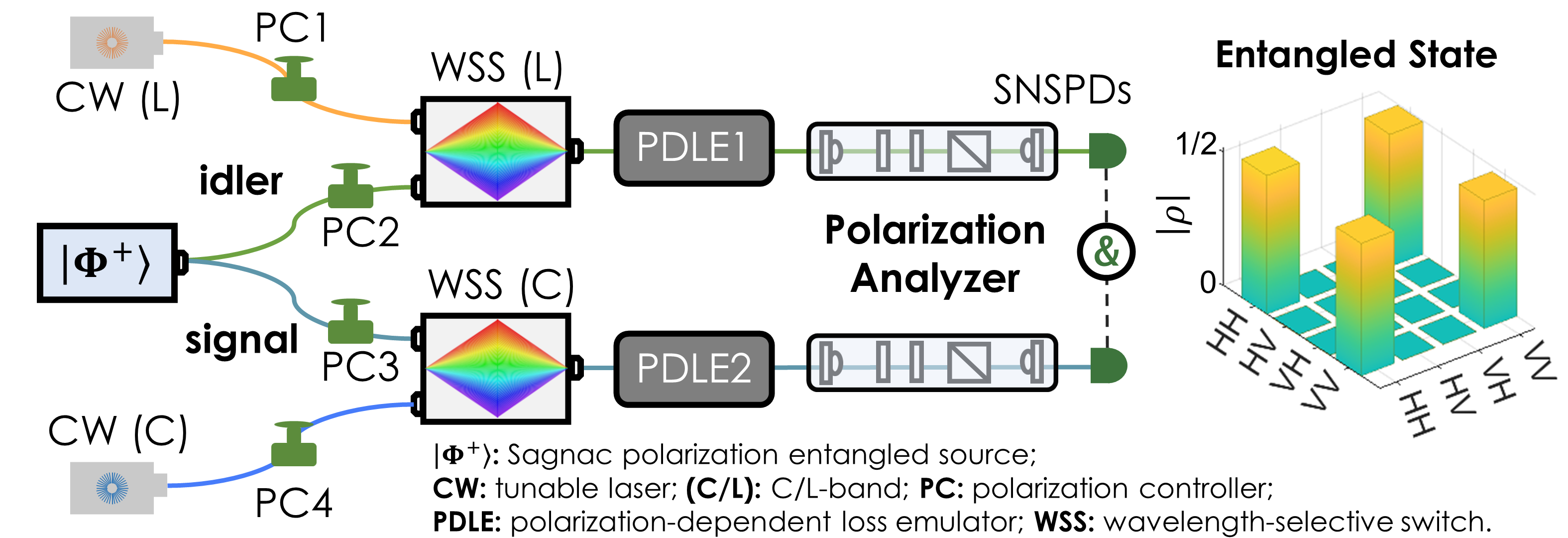}
\caption{Experimental setup and Bayesian mean density matrix of the initial entangled resource. See text for details.} 
\label{fig3}
\end{figure}

\textit{Tabletop experiments.---}Figure~\ref{fig3} illustrates the experimental setup. Broadband, polarization-entangled photons are
separated by a C/L-band demultiplexer, with signal and idler photons covering the optical C-band (1530--1565 nm) and L-band (1565--1625 nm), respectively [detailed design outlined in Ref.~\cite{Alshowkan2022b, Lu2023b}]. In each fiber arm, we program a wavelength-selective switch (WSS; Finisar) operated in reverse to accept a 25~GHz-wide energy-matched channel pair: 193.9~THz in the C-band and 189.1~THz in the L-band. 
For all scenarios tested, we measure coincidences (5~s per point in the tabletop experiments, 10~s in the deployed tests) in an overcomplete set of 36 polarization projections using a pair of motor-controlled polarization analyzers~\cite{Alshowkan2022b, Lu2023b}. 
Bayesian quantum state tomography (QST)~\cite{BlumeKohout2010, Lukens2020b}, following the refinements 
outlined in Ref.~\cite{Lu2022}, is then applied to recover density matrix samples. Without any crosstalk and a coincidence window $\tau=1.4$~s, we find the mean density matrix in Fig.~\ref{fig3}, with fidelity $\sF=0.988(1)$.

Subsequently, we introduce classical crosstalk using a pair of tunable, linearly polarized continuous-wave (CW) lasers, 
operating at frequencies of 189.125 THz [CW (L) in Fig.~\ref{fig3}] and 193.925 THz [CW (C) in Fig.~\ref{fig3}], respectively, i.e., one frequency slot away from their respective quantum channel on the 25~GHz ITU grid (ITU-T Rec. G.694.1). To emulate the narrowband filtering necessary for future coexistence networks
, both lasers are heavily attenuated. 
At each receiver, we record $\nu \approx 4.5\times 10^{5}$ s$^{-1}$, chosen to introduce an aggressive amount of accidentals while avoiding saturation of the superconducting nanowire single-photon detectors (SNSPDs)---the regime of interest for our Procrustean method.

\begin{figure*}[bt!]
\includegraphics[width=7in]{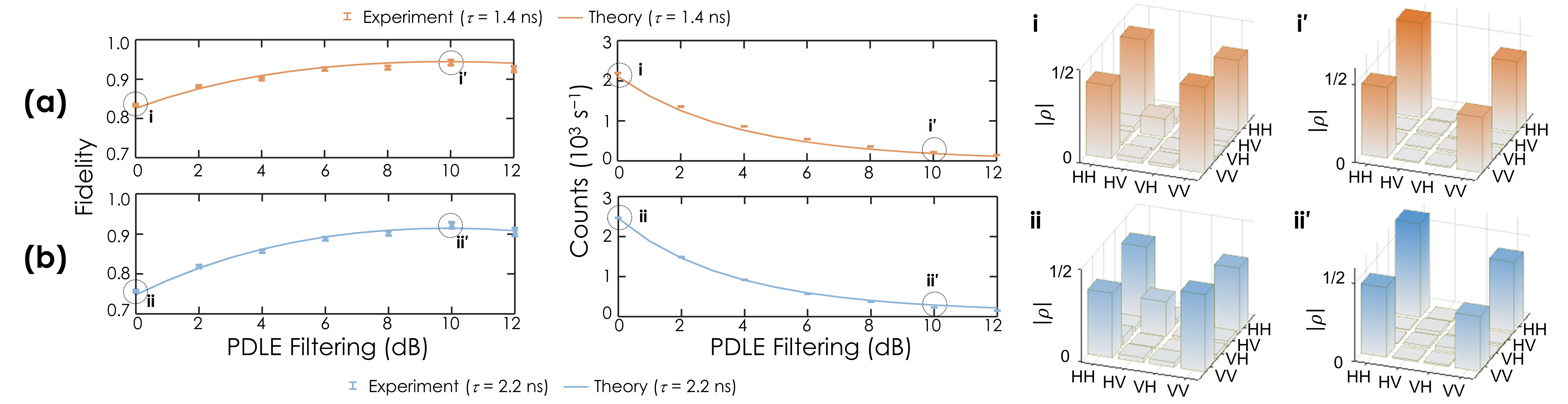}
\caption{Bayesian state fidelities and collected coincidences versus PDLE local filtering.  Measured density matrices at (i,ii) 0 dB ($T_H = T_V = 1$) and (i$'$,ii$'$) 10 dB ($T_H = T_V = 0.1$) of PDL . Coincidence window $\tau$: (a) 1.4 ns, (b) 2.2~ns.}
\label{fig4}
\end{figure*}

To demonstrate the method for maximum improvement relies on two levels of coordination in polarization: first, the PDL axis at each receiver must be aligned to the polarization of the classical crosstalk; second, the polarization correlations for the crosstalk contributions must be orthogonal to the correlations in the Bell state of interest. Once the former is achieved, the latter can in principle be realized through local rotation of only one photon from the maximally entangled pair (prior to multiplexing)~\cite{Wilde2017, Valencia2020}. Nonetheless, for ease of implementation here, we manually tune the polarization of all four inputs relative to fixed PDL axes [via polarization controllers (PCs) 1--4 in Fig.~\ref{fig3}]; 
this allows us to produce a state in the form of Eq.~(\ref{eq:theory}), up to a relative phase between $\ket{HH}$ and $\ket{VV}$ in the Bell state term. Any residual phase is automatically compensated by the local rotations we apply to the inferred density matrices in post-processing, chosen to maximize fidelity with respect to $\ket{\Phi^+}$~\cite{Alshowkan2022b}. The resulting noisy state measured with $\tau=$1.4 ns is depicted in Fig.~\ref{fig4}(a) (labeled i) and resembles a MEMS~I state~\cite{Munro2001, Wei2003, Peters2004a}.

For the above and all experimental results below, we estimate the output density matrix $\rho$ and flux $R$ with no reference to Eq.~(\ref{eq:theory}); the states are inferred from a completely uniform Bures prior without assuming any particular noise model~\cite{Lu2022}. However, to extract critical parameters of interest for validating our model, for this particular case only ($T_H=T_V=1$ and $\tau=1.4$~ns) we perform an additional inference step in which we temporarily assume Eq.~(\ref{eq:theory}) to infer the parameters $(\mu,\eta_A,\eta_B,\nu_A,\nu_B)$, which are difficult to measure independently. For this purpose, we develop a customized Bayesian model patterned after Ref.~\cite{Lu2019a} that considers both single-detector events 
and coincidences, 
takes as input the independently measured dark count rates at $d_A=d_B=100$~s$^{-1}$, and places normal priors on the four unknowns of interest $(\mu,\eta_A,\eta_B,\nu_A,\nu_B)$, with means given by our initial estimates from the count data, and standard deviations equal to 10\% of these values. Under this ``open box'' model~\cite{Lu2019a, Lu2023a}, we obtain $(\mu,\eta_A,\eta_B,\nu_A,\nu_B) = (5.889(3)\times10^{6}\,\mathrm{s}^{-1}, 2.113(1)\times10^{-2}, 1.375(2)\times10^{-2}, 4.152(1)\times10^{5}\,\mathrm{s}^{-1}, 4.653(2)\times10^5\,\mathrm{s}^{-1})$,
which, in conjunction with Eq.~(\ref{eq:theory}), serve as the basis for subsequent theory curves. 

Figure~\ref{fig4}(a) depicts the state evolution as the PDLE filtering level is increased from 0~dB ($T=1$) to 12~dB ($T = 0.063$) in 2~dB steps. The fidelity advances steadily from $\sF_{0~\mathrm{dB}}=83.4(3)\%$ to $\sF_{10~\mathrm{dB}}=94.2(7)\%$ [Fig.~\ref{fig4}(a)], at the expense of reduced flux [Fig.~\ref{fig4}(b)]. Incidentally, the linear entropy decreases from $0.384(5)$ to $0.10(2)$, while the tangle increases from $0.57(1)$ to $0.81(3)$, confirming true entanglement ``concentration'' as defined as increases in both entanglement and purity~\cite{Thew2001, Peters2004a}. We halt the investigation at $12$~dB as we have observed that, for our particular system parameters, the fidelity reaches a plateau within the range of 10--12 dB, as predicted by the accompanying theory. 

To examine noisier state, we widen the coincidence window to $\tau=$2.2~ns, resulting in a roughly 60\% increase in accidentals [cf. Eq.~(\ref{eq:theory})]. The results, depicted in Fig.~\ref{fig4}(b), show Procrustean fidelity improvements from $\sF_{0~\mathrm{dB}}=75.8(3)\%$ to $\sF_{10~\mathrm{dB}}=92.2(8)\%$. Additionally, the overall trend and optimal filtering level closely match those observed in the case of Fig.~\ref{fig4}(a) when $\tau=$1.4 ns, which aligns with our findings in Fig.~\ref{fig2}(b).

\emph{Deployed network tests.---}We next apply our method in a deployed quantum local area network on the Oak Ridge National Laboratory campus~\cite{Alshowkan2021, Alshowkan2022a}.
The source and components in the idler arm remain in the original lab (Alice in \cite{Alshowkan2021, Alshowkan2022a}), while PDLE2 and the polarization analyzer in the signal arm are relocated to another building (Bob in~\cite{Alshowkan2021, Alshowkan2022a}). A deployed fiber link, approximately 250~m long with a round-trip loss of $\sim$5 dB, connects the two buildings. After polarization projections, signal photons are routed back to Alice for photon detection. We transmit Bob's projected photons back to Alice---rather than detect them at Bob---due to the availability of SNSPDs; Bob's avalanche photodiodes considered in previous experiments~\cite{Alshowkan2021, Alshowkan2022a} approach saturation at the level of crosstalk noise of interest here.

\begin{figure}[b!]
\includegraphics[width=\columnwidth, trim=0.5cm 0cm 0.5cm 0cm,clip]{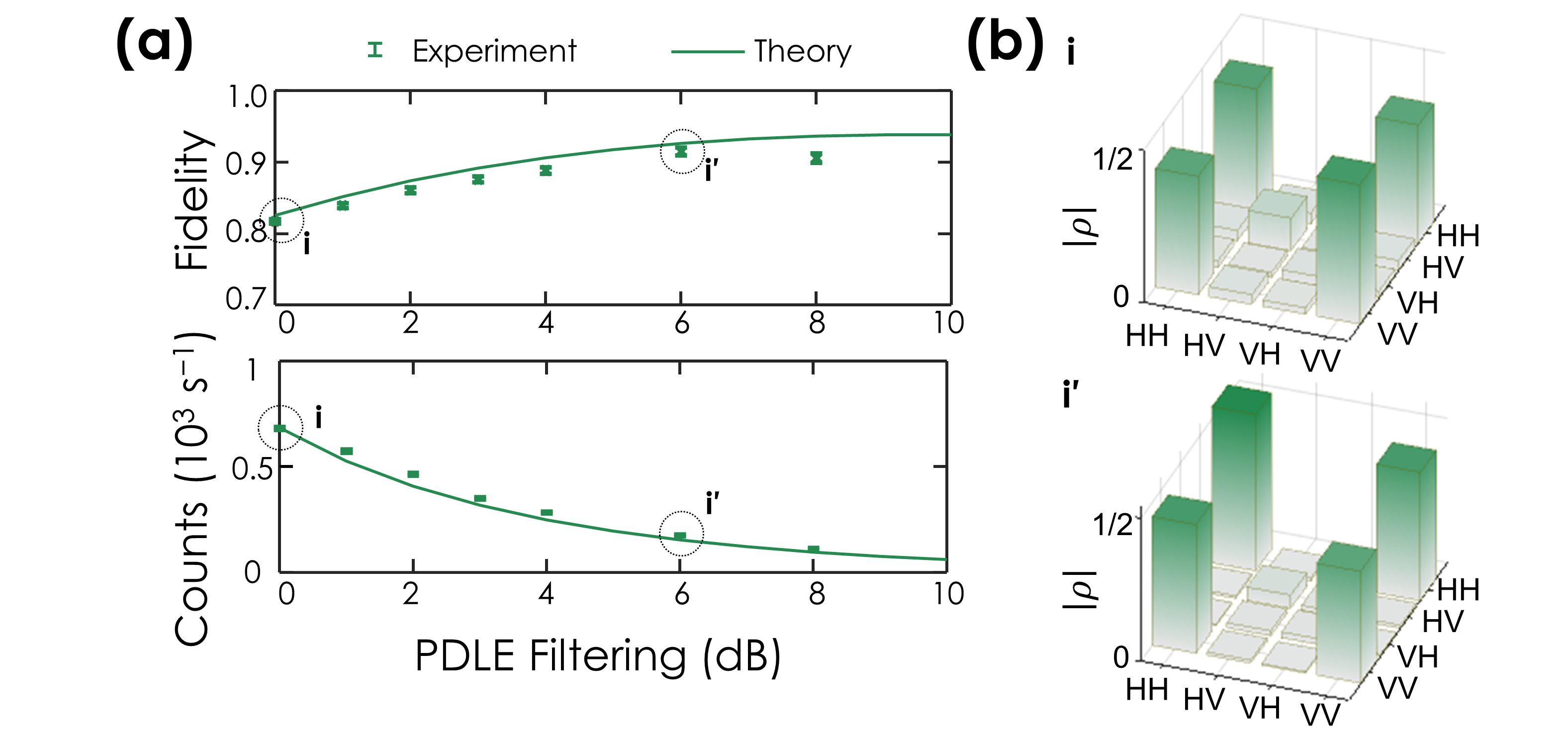}
\caption{(a) Bayesian-estimated fidelities and flux versus PDLE local filtering in a deployed network setting.  (b) Measured density matrices at (i) 0 dB and (i$'$) 6~dB PDL.}
\label{fig5}
\end{figure}

As a further modification from the tabletop experiment, we tune the classical crosstalk lasers from their original positions 
to the center of their respective quantum channels at 189.1~THz and 193.1~THz. Now, the Procrustean filtering approach described here places no requirements on the specific wavelength of the crosstalk, as long as it is polarized and copropagating with the demultiplexed quantum output. Thus, whether stemming from imperfect isolation of an nearby frequency (emulated in Fig.~\ref{fig4}) or from background light that truly spectrally overlaps with the quantum signal (emulated in Fig.~\ref{fig5}), all that matters is its overall polarized contribution to the total rate of received photons. 

Figure~\ref{fig5} summarizes the experimental results obtained from our deployed network. In this case, Procrustean filtering improves the entangled state resource from 
$\sF_{0~\mathrm{dB}}=81.7(3)\%$ to $\sF_{6~\mathrm{dB}}=91.5(5)\%$, which so happens to transition the distributed entangled state from a regime where quantum key distribution (QKD) is impossible to one where it may be performed. The BB84 QKD protocol using one-way communication possesses an 11\% quantum bit error rate (QBER) threshold~\cite{Shor2000, Koashi2003, Ma2007}; the QBERs associated with the initial state are 13.8(4)\% and 12.2(4)\% for the rectilinear and diagonal bases, respectively, but they decrease to 6.6(6)\% and 3.8(6)\% for the 6~dB filtered state.

\textit{Discussion.---}In this study, we have employed fidelity as the key metric to assess the effectiveness of Procrustean filtering. However, our model can be adapted to optimize other metrics that are computable from the density matrix as well~\cite{Peters2004b, Vidal2002, Plenio2005, Devetak2005}. 
Expanding our analysis, we could employ constrained nonlinear optimization to explore more complex scenarios, such as maximizing total coincidence rates while constraining fidelity $\sF$ above a predefined application-specific threshold $\sF_\mathrm{th}$~\cite{Alnas2022, Gayane2022}. Such an approach can be particularly valuable in photon-starved applications 
where optimizing both throughput and quality of entanglement are essential.

Procrustean filtering should not be construed as an alternative to CWDM/DWDM but rather an additional layer that can supplement such strategies to further mitigate the impact of classical crosstalk in shared optical fiber. By coordinating the relative orientations of entangled photons and coexisting classic traffic, the two-photon correlations possessed by the quantum and classical portions are distinct and can be selectively filtered when multiplexed in shared fiber. 
Accordingly, while our procedure does presuppose polarization coordination between coexisting quantum and classical signals, the benefits it engenders are quite general and apply to a variety of quantum--classical operating conditions in which the dominant crosstalk is polarized. Such cases can lead to asymmetric errors in the quantum state, making this local filtering technique a potentially valuable preliminary step, in tandem with other error reduction techniques, for improving entanglement distribution efficiency in 
future quantum repeater networks.

\begin{acknowledgments}
We thank Y. Zhang for helpful discussions regarding QBER thresholds. This work was performed in part at Oak Ridge National Laboratory, operated by UT-Battelle for the U.S. Department of energy under contract no. DE-AC05-00OR22725. Funding was provided by the U.S. Department of Energy, Office of Science, Advanced Scientific Computing Research (Field work Proposals ERKJ381, ERKJ353).
\end{acknowledgments}

%

\end{document}